\documentclass[12pt]{article}
\pdfoutput=1

\usepackage{xcolor}
\definecolor{darkblue}{rgb}{0.1,0.1,.7}
\usepackage[colorlinks, linkcolor=darkblue, citecolor=darkblue, urlcolor=darkblue, linktocpage,pagebackref]{hyperref} 
\usepackage[english]{babel}
\usepackage{amsmath,amssymb,amsthm,graphicx}
\usepackage{booktabs}
\usepackage{geometry}
\usepackage[margin=10pt,font=small,labelfont=bf]{caption}

\usepackage{titlesec}
\titleformat*{\section}{\large\bfseries}
\titleformat*{\subsection}{\normalsize\bfseries}
\titleformat*{\subsubsection}{\normalsize\it}
\titleformat*{\paragraph}{\normalsize\bfseries}
\titleformat*{\subparagraph}{\normalsize\bfseries}

\newtheorem{mydef}{Definition}
\newtheorem{axiom}{Axiom}
\newtheorem{myremark}{Remark}
\usepackage[utf8]{inputenc}

\begin{document}

\vspace*{-.6in} \thispagestyle{empty}
\begin{flushright}
\end{flushright}
\vspace{.2in} {\Large
	\begin{center}
		\resizebox{\textwidth}{!}{\bf  3D Ising Model: a view from the Conformal Bootstrap Island}\\
	\end{center}
}
\vspace{.2in}
\begin{center}
	{\bf 
		Slava Rychkov$^{a,b}$
	} 
	\\
	\vspace{.2in} 
	{\it $^{a}$ Institut des Hautes \'Etudes Scientifiques, Bures-sur-Yvette, France}\\
  {\it $^{b}$ Laboratoire de Physique de l’Ecole normale sup\'erieure, ENS, \\
		Universit\'e PSL, CNRS, Sorbonne Universit\'e, Universit\'e de Paris, F-75005 Paris, France}
\end{center}

\vspace{.2in}

\begin{abstract}
  We explain how the axioms of Conformal Field Theory are used to
	make predictions about critical exponents of continuous phase transitions in
	three dimensions, via a procedure called the conformal bootstrap. The method assumes conformal invariance of correlation functions, and imposes some relations between correlation functions of different orders. Numerical analysis shows that these conditions are incompatible unless the critical exponents take particular values, or more precisely that they must belong to a small island in the parameter space.
\end{abstract}

\section{Introduction}

Physics has many \textit{emergent laws}, which follow in a non-obvious way
from more fundamental microscopic laws. Whenever this happens, we have two
separate goals: to understand how the emergent law arises, and to explore its
consequences.

One example is the Gibbs distribution of equilibrium statistical mechanics:
the probability for the system in thermal equilibrium at temperature $T$ to be
found in a state $n$ of energy $E_n$ is proportional to $\exp (- E_n / T)$.
One may be interested in deriving this emergent law from microscopic models of
thermalization, or in exploring the myriad of its physical consequences.

This text, based on a recent talk for an audience of mathematical physicists,
is about the ``conformal field theory'' (CFT), a set of emergent laws
governing critical phenomena in equilibrium statistical mechanics (such as the
liquid-vapor critical point or the Curie point of ferromagnets). CFT makes
certain assumptions about the state of the system at a critical point. These
assumptions can be given reasonable physical explanations, but for the
purposes of this talk we will view them as axioms.

CFT is an ``emergent law of second degree'' with respect to the Gibbs
distribution, by itself emergent. It sidesteps the Gibbs distribution
similarly to how the Gibbs distribution sidesteps a thermalization model. Future work should derive the CFT axioms with mathematical
rigor from the Gibbs distribution. Our goal here will be to explain the
axioms and how they lead to concrete predictions for observable quantities
through a procedure called the ``conformal bootstrap''.

CFT/bootstrap approach to critical phenomena is an alternative to the
better-known Wilson's renormalization group (RG) theory.
The RG is more directly
related to the Gibbs distribution than CFT, although it too is not fully
mathematically justified. The RG will not be treated here except for a few
comments.

We will not give many references, which can be found in the recent review
{\cite{Poland:2018epd}}. See also lecture notes {\cite{Rychkov:2016iqz,Simmons-Duffin:2016gjk,Chester:2019wfx}}. An excellent set of
recorded lectures is {\cite{Petr}}.

\section{The first two CFT axioms}\label{A1A2}

We will describe the axioms of Conformal Field Theory (CFT) on $\mathbb{R}^d$,
$d \geqslant 3$. These axioms are well established in the physics literature.
We will present them in a form hopefully more accessible to mathematicians. In
particular, we will try to avoid (or at least explain) excessive physics
jargon. Similar axioms, with additional bells and whistles, hold in $d = 2$
dimensions \cite{DiFrancesco:1997nk}.\footnote{Our axioms should be viewed as a sketch of future
complete axiomatics, which has not yet been written up in
the mathematics literature. A different approach to axiomatize CFTs in $d \geqslant 3$ (akin to Segal's
2d CFT axiom) is in \cite{Schwarz:2015fva}, but it makes the connection to
concrete calculations less explicit. A nicer starting point is the recent mathematics paper {\cite{Moriwaki:2020dlj}} which develops 2d CFT from the conformal bootstrap perspective. It would be interesting to generalize it to $d \geqslant 3$. }

\textit{Suppose we are given a collection of real-valued functions
\begin{eqnarray}
  \mathcal{T}= \{ G_{i_1, \ldots, i_n} (x_1, \ldots , x_n) \}, \label{TG}
\end{eqnarray}
defined for $x_p\in\mathbb{R}^d$, $x_p\ne x_q$ $(p,q=1,\ldots,n)$, where $n \geqslant 1$ and the indices $i_p$ are non-negative
integers.}

Functions \eqref{TG} are called ``$n$-point correlators of fields
$A_{i_1}$, {...}, $A_{i_n}$'' and are also denoted by
\begin{equation}
  \langle A_{i_1} (x_1) A_{i_2} (x_2) \ldots A_{i_n} (x_n) \rangle .
  \label{notG}
\end{equation}
The collection $\mathcal{T}$ is called a CFT if it satisfies
certain axioms stated below. Different CFT's are just different collections of
correlators satisfying those axioms.\footnote{One also uses the term
``Conformal Field Theory'' in a meta-sense, as the study of all possible
CFTs.}

Note that the ``field'' $A_i$ is just a label, a name, and Eq.~{\eqref{notG}}
is just a notation for $G_{i_1, \ldots, i_n} (x_1, \ldots ., x_n)$. The
statistical average operation suggested by this notation does not have a
direct meaning in the CFT axioms. It will be handy in the
\textit{interpretation} of the axioms (section \ref{interpretation}).

\begin{axiom}[Simple properties]
  \label{AxSimple}Correlators have the following properties:
  \begin{enumerate}
    \item[(a)] They are invariant under permutation of any two fields:\footnote{One
    can also consider CFTs with fields having fermionic statistics, whose
    correlators change sign under permutations. Such CFTs are
    important e.g. for describing quantum critical points of many-electron
    systems. Here we only consider commuting fields for simplicity.}
    \begin{equation}
      \langle A_{i_1} (x_{}) A_{i_2} (y) \ldots \rangle = \langle A_{i_2} (y)
      A_{i_1} (x_{}) \ldots \rangle, \text{etc} .
    \end{equation}
    \item[(b)] Index $i = 0$ is associated with the ``unit field'', 
    replaced by 1 under the correlator sign:
    \begin{equation}
      \langle A_0 (x) \times \text{anything} \rangle = \langle \text{anything}
      \rangle .
    \end{equation}
    \item[(c)] The 1-point (1pt) correlators are given by
    \begin{equation}
      \langle A_0 (x) \rangle \equiv 1, \qquad \langle A_i (x) \rangle \equiv
      0 \qquad (i \geqslant 1) .
    \end{equation}
    \item[(d)] The 2pt correlators are given by
    \begin{equation}
      \langle A_i (x) A_j (y) \rangle = \frac{\delta_{i j}}{| x - y |^{2
      \Delta_i}}  \quad (x, y \in \mathbb{R}^d) . \label{2pt}
    \end{equation}
    where $\delta_{i j}$ is the Kronecker symbol, and $\Delta_i \geqslant
    \frac{d - 2}{2}$ ($i \geqslant 1$) is a real number called ``scaling
    dimension of field $A_i$''. For the unit field we have $\Delta_0 = 0$.
    
    \item[(e)] The set of scaling dimensions $\{ \Delta_i \}$ is called the
    ``spectrum''. It is a discrete set without accumulation points (i.e. there
    are finitely many scaling dimensions below any $\Delta_{\ast} < \infty$).
  \end{enumerate}
\end{axiom}

\begin{axiom}[Conformal invariance]
  \label{AxConf}Correlators are conformally invariant, in the sense
  that they satisfy the constraint
  \begin{equation}
    G_{i_1, \ldots, i_n} (x_1, \ldots ., x_n) = \left( \prod_{p = 1}^n \lambda
    (x_p)^{\Delta_{i_p}} \right) G_{i_1, \ldots, i_n} (f (x_1), \ldots ., f
    (x_n)), \label{confinv}
  \end{equation}
  or equivalently, using notation {\eqref{notG}},
  \begin{equation}
    \langle A_{i_1} (x_1) \ldots A_{i_n} (x_n) \rangle = \left( \prod_{p =
    1}^n \lambda (x_p)^{\Delta_{i_p}} \right) \langle A_{i_1} (f (x_1)) \ldots
    A_{i_n} (f (x_n)) \rangle,
  \end{equation}
  where $f (x)$ is an arbitrary conformal transformation of $\mathbb{R}^d$ and $\lambda (x) = \left| \frac{\partial f}{\partial x}
  \right|^{1 / d}$ is its scale factor.
\end{axiom}

Recall that conformal transformations satisfy the constraint $\partial^{}
f^{\mu} / \partial x^{\nu} = \lambda (x) R^{\mu}{}_{\nu} (x)$ where
$R^{\mu}{}_{\nu} (x) \in \text{SO} (d)$. For $d \geqslant 3$, these transformations form a group $\text{SO} (d + 1, 1)$.

\begin{myremark} \rm
	Conformal transformations of $\mathbb{R}^d$ may send points to infinity, and should be thought more properly as acting on $\mathbb{R}^d \cup \{\infty \}$, the $d$-dimensional analogue of the Riemann sphere.
To treat the point at infinity on equal footing with the other points, one can put $\mathbb{R}^d \cup \{\infty \}$ in one-to-one correspondence with the $d$-dimensional unit sphere $S^d$ via the stereographic projection.
This subtlety will be glossed over here.
\end{myremark}
%allows to give a more careful formulation of Axiom \eqref{AxConf}.

\subsection{Basic consequences of conformal invariance}

We will state without proof a few basic consequences of the above axioms. One
can check that the 2pt correlators given in Axiom \ref{AxSimple}(d)
are consistent with Axiom \ref{AxConf}. Note that the same scaling dimension
$\Delta_i$ has to appear in all $n$-point correlators involving the field $A_i$.
The $3$pt correlators are fixed by Axiom \ref{AxConf} up to an overall factor:
\begin{equation}
  \langle A_i (x_1) A_j (x_2) A_k (x_3) \rangle = \frac{c_{i j k}}{x^{\Delta_i
  + \Delta_j - \Delta_k}_{12} x_{13}^{\Delta_i + \Delta_k - \Delta_j}
  x_{23}^{\Delta_j + \Delta_k - \Delta_i}}, \label{3pt}
\end{equation}
where $c_{i j k}$ is totally symmetric by Axiom \ref{AxSimple}(a), and we
denoted $x_{i j} = | x_i - x_j |$. For 4pt correlators Axiom \ref{AxConf}
implies the following functional form:
\begin{equation}
  \langle A_i (x_1) A_j (x_2) A_k (x_3) A_l (x_4) \rangle = \left(
  \frac{x_{24}}{x_{14}} \right)^{\Delta_i - \Delta_j} \left(
  \frac{x_{14}}{x_{13}} \right)^{\Delta_k - \Delta_l} \frac{g_{i j k l} (u,
  v)}{x_{12}^{\Delta_i + \Delta_j} x_{34}^{\Delta_k + \Delta_l}}, \label{4pt}
\end{equation}
where $g_{i j k l} (u, v)$ is a function of conformally invariant
cross-ratios:
\begin{equation}
  u = \frac{x_{12}^2 x_{34}^2}{x_{13^{}}^2 x_{24}^2}, \qquad v = u |_{1
  \leftrightarrow 3}  = \frac{x_{23}^2 x_{14}^2}{x_{13^{}}^2
  x_{24}^2} .
\end{equation}
By Axiom \ref{AxSimple}(a) functions $g_{i j k l}$ with permuted indices are
all related, e.g. permutation $1 \leftrightarrow 3$ generates the constraint:
\begin{equation}
  u^{- \frac{\Delta_i + \Delta_j}{2}} g_{i j k l} (u, v) = v^{- \frac{\Delta_k
  + \Delta_j}{2}} g_{k j i l} (v, u), \text{etc} . \label{4ptcross}
\end{equation}

\subsection{Primaries and descendants}

Group-theoretically, the transformation
\begin{equation}
  A (x) \rightarrow \lambda (x)^{\Delta} A (f (x)) \label{scalar}
\end{equation}
is an irreducible representation $\pi_{\Delta}$ of the conformal group on
scalar functions $A : \mathbb{R}^d \rightarrow \mathbb{R}$.
Eq.~{\eqref{confinv}} means that the correlators $G_{i_1, \ldots,
i_n}$ belong to the invariant subspace of the tensor product representation
$\otimes_{p = 1}^n \pi_{\Delta_{i_p}}$ (so they can be called ``invariant
tensors'').

We formulated Axioms 1,2,3 for the fields transforming as {\eqref{scalar}},
called ``scalar fields''. These axioms can and should be extended to allow for
fields with tensor indices. First of all, we have to add fields
$\partial_{}^{\alpha} A_{i_{}} (x_{})$ which are partial derivatives (of
arbitrary order) of the fields $A_{i_{}}$. Their correlators are defined as
derivatives of the original ones:
\begin{equation}
  \langle \partial_{}^{\alpha} A_{i_{}} (x_{}) \ldots \rangle :=
  \partial^{\alpha}_x \langle A_{i_{}} (x_{}) \ldots \rangle . \label{dA}
\end{equation}
This is, in a sense, just a convenient notation. The basic fields $A_i (x)$
whose correlators transform as {\eqref{scalar}} are called ``primaries'',
while their derivatives ``descendants''. Transformation rules for correlators
of descendants can be obtained by differentiating {\eqref{scalar}}.

The second extension is a bit less trivial. We should generalize {\eqref{scalar}},
allowing for fields with values in a finite-dimensional vector space $V$,
$\dim V > 1$, transforming under the conformal group via
\begin{equation}
  A (x) \rightarrow \lambda (x)^{\Delta} \rho (R (x)) A (f (x)),
  \label{spinning}
\end{equation}
where $\rho$ is an irreducible representation of $\text{SO} (d)$ acting in
$V$. Such fields are called ``primary spinning fields''. One example is $V =
\left\{ \text{symmetric traceless rank-$l$ tensors} \right\}$. Correlators of spinning fields then take values in the tensor product
$\otimes_{p = 1}^n V_{i_p}$ and satisfy a conformal invariance constraint
similar to {\eqref{confinv}} but with factors of $\rho_{i_p} (R (x_p))$ in the
l.h.s. [Derivatives of spinning fields are then also added as in
{\eqref{dA}}.] Adding spinning fields would complicate the notation a bit. We will neglect them here, 
although practical conformal bootstrap computations always allow
for their presence.

\section{The OPE axiom}\label{sec:AxOPE}

The last ``OPE axiom'' will relate different correlators, and in particular
correlators with different $n$. This is unlike the previous axioms which
involved one $n$-point correlator at a time.\footnote{Except the rather
trivial Axiom \ref{AxSimple}(b).} \

Suppose we are given two collections of real numbers
\begin{equation}
  \{ \lambda_{i j k} \}, \{ s_{i j k}^{(r)}, r \geqslant 1 \},
  \label{OPEcoeff}
\end{equation}
where $i, j, k$ run over the field indices (non-negative integers). With these
numbers as coefficients, \ ``Operator Product Expansion'' (OPE) is constructed
as a set of formal equalities (one for each pair of fields $A_i$ and $A_j$):
\begin{eqnarray}
  A_i (x) A_j (y) & = & \sum_{k = 0}^{\infty} \frac{\lambda_{i j k}}{| u
  |^{\Delta_i + \Delta_j - \Delta_k}} \nonumber\\
  &  & \times \left[ A_k (x) + s_{i j k}^{(1)} u^{}_{\mu} \partial^{\mu}_x
  A_k (x) + \left( s_{i j k}^{(2)} u^{}_{\mu} u^{}_{\nu} + s_{i j k}^{(3)} u^2
  \delta_{\mu \nu} \right) \partial^{\mu}_x \partial^{\nu}_x A_k (x) + \cdots
  \right] . 
\end{eqnarray}
where $u = y - x$. Using the OPE for the first pair of fields inside the
$n$-point correlator {\eqref{notG}} with $n \geqslant 2$, we get a set of
candidate identities among correlators:
\begin{eqnarray}
  \langle A_{i_{}} (x_{}) A_j (y_{}) \boldsymbol{\Pi} \rangle & = & \sum_{k =
  0}^{\infty} \frac{\lambda_{i j k}}{| u |^{\Delta_i + \Delta_j - \Delta_k}}
  [\langle A_k (x) \boldsymbol{\Pi} \rangle + s_{i j k}^{(1)} u^{}_{\mu}
  \partial^{\mu}_x \langle A_k (x) \boldsymbol{\Pi} \rangle + \cdots], 
  \label{OPEeq}
\end{eqnarray}
where we denoted $i_1 = i$, $i_2 = j$, $x_1 = x$, $x_2 = y$, and
$\boldsymbol{\Pi}= \Pi_{p = 3}^n A_{i_p} (x_p)_{}$ is the product of all other
fields in the correlator. In the l.h.s. we have an $n$-point correlator, while
in the r.h.s. we have an infinite series of $(n - 1)$-point correlators and
derivatives thereof.

The OPE axiom gives a condition for when the candidate identity
{\eqref{OPEeq}} is a true identity.

\begin{figure}[h]
	\centering
	\includegraphics[width=0.3\linewidth]{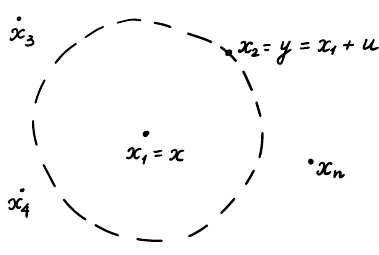}
  \caption{\label{OPEcond}The OPE expansion applies when all points $x_p$
  ($p \geqslant 3$) lie further from $x_1$ than $x_2$.}
\end{figure}

\begin{axiom}[OPE]
  \label{AxOPE}There exists a set of coefficients {\eqref{OPEcoeff}}, such
  that Eq.~{\eqref{OPEeq}} holds as a true relation between correlators (the series in the r.h.s. converges absolutely to the l.h.s.) as
  long as $| x_p - x | > | u |$ for all $p \geqslant 3$ (see Fig.
  \ref{OPEcond}).
\end{axiom}

What can be said about coefficients {\eqref{OPEcoeff}} which make this axiom
work? To see this, \ let us apply Eq.~{\eqref{OPEeq}} to a 3pt correlator.
Because the 2pt correlators vanish for non-identical fields ($\delta_{i j}$ in
Axiom \ref{AxSimple}(d)), the sum in the r.h.s. collapses to the single $k$
value, and we get:
\begin{equation}
  \langle A_{i_{}} (x_{}) A_j (y_{}) A_k (x_3) \rangle = \frac{\lambda_{i j
  k}}{| u |^{\Delta_i + \Delta_j - \Delta_k}} \left[ \frac{1}{| x - x_3 |^{2
  \Delta_k}} + s_{i j k}^{(1)} u^{}_{\mu} \partial^{\mu}_x \frac{1}{| x - x_3
  |^{2 \Delta_k}} + \cdots \right], \label{OPEn3}
\end{equation}
where we used that $\langle A_k (x) A_k (x_3) \rangle = 1 / | x - x_3 |^{2
\Delta_k}$. On the other hand, we already know that the $3$pt correlator in the
l.h.s. has form {\eqref{3pt}} by Axiom \ref{AxConf}. Let us then expand
{\eqref{3pt}} for small $y$ and match with {\eqref{OPEn3}}. From the leading
term we find $\lambda_{i j k} = c_{i j k}$.\footnote{\label{lNote}In
particular we learn that $\lambda_{i j k}$ has to be symmetric, just as $c_{i
j k}$. Note also that by putting $A_k = A_0 = 1$ in {\eqref{3pt}} and by using
Axiom \ref{AxSimple}(d), we get $\lambda_{i j 0} = c_{i j 0} = \delta_{i j}$.}
Relative to this overall normalization, the subleading terms on the l.h.s. are
fixed, and this allows to determine $s_{i j k}^{(r)}$ uniquely as rational
functions of $\Delta_i$, $\Delta_j$, $\Delta_k$ and $d$. We conclude that all
coefficients {\eqref{OPEcoeff}} can be uniquely determined by demanding that
the OPE axiom works for the 3pt correlators. Furthermore, the axiom says that
the same set of coefficients should then also work for any $n$-point correlators.

\begin{mydef}
  A conformal field theory (CFT) $\mathcal{T}$ in $d \geqslant 3$ dimensions
  is a collection of correlators {\eqref{TG}} satisfying Axioms
  \ref{AxSimple}-\ref{AxOPE}. 
\end{mydef}

\begin{myremark} \rm
  We included in our axioms the conditions $\Delta_i \geqslant \frac{d -
  2}{2}$ for $i \geqslant 1$ in Axiom $\ref{AxSimple}$(d), and $\lambda_{i j
  k} \in \mathbb{R}$ in Eq.~{\eqref{OPEcoeff}}. By this, we are restricting
  our discussion to a subclass of conformal field theories called reflection
  positive (or unitary). Many statistical physics systems at criticality (such
  as the Ising model or $O (N)$ models) are known to be described by unitary
  CFTs. %In a full version of CFT axioms, including spinning fields and
  %fermions, reflection positivity will have to be given a more careful treatment.
\end{myremark}

\subsection{CFT data}\label{CFTdata}

The spectrum $\Delta_i $ and the OPE coefficients $\lambda_{i
j k}$ comprise the ``dataset'' of a CFT $\mathcal{T}$:
\begin{equation}
  \text{Data} (\mathcal{T}) = \{ \Delta_i, \lambda_{i j k} \} .
\end{equation}
As discussed, $\text{Data} (\mathcal{T})$ is in one-to-one correspondence with
the 2pt and 3pt correlators of $\mathcal{T}$. Moreover, knowing $\text{Data}
(\mathcal{T})$ we can reconstruct all $n$-point correlators, for an arbitrarily high $n$.
Indeed, from $\text{Data} (\mathcal{T})$ we can construct the OPE (the
coefficients $s_{i j k}^{(r)}$ are not included in $\text{Data} (\mathcal{T})$
since they are uniquely determined by $\Delta_{}$'s and $d$). Then, we can recursively reduce any $n$-point correlator to lower-point ones,
until we get to the known 2pt and 3pt correlators.\footnote{We should take care
that the OPE is used for a pair of fields at positions $x_1, x_2$ verifying
conditions of Axiom \ref{AxOPE}, so that it converges. This is the case if
$x_2$ is the unique position with the minimal distance from $x_1$. There are
degenerate configurations when such a pair cannot be found, because each point
has two or more nearest neighbors at equal distance (e.g. the vertices of a
regular polygon). It is then always possible to apply a small conformal
transformation which moves points to a non-degenerate configuration. In the new configuration the OPE
converges and we can compute the value of the correlator. We then
conformal-transform back to the original configuration. This way we can
compute correlators in any configuration of non-coincident points.}

We thus see that the dataset $\text{Data} (\mathcal{T})$ encodes full
information about the CFT $\mathcal{T}$. Below we will describe a program of
classifying CFTs by classifying their data sets. But first let us
discuss the interpretation of the CFT axioms.

\section{Interpretation}\label{interpretation}

Notation $\langle A_{i_1} (x_1) A_{i_2} (x_2) \ldots A_{i_n}
(x_n) \rangle$ for $G_{i_1, \ldots, i_n} (x_1, \ldots ,
x_n)$ acquires a meaning in the interpretation of the CFT
axioms, as correlation functions of statistical systems at their critical
points. CFT calculations are then interpreted as predictions for the critical exponents of statistical physics models. Although the CFT calculations based on the axioms are completely rigorous, the interpretation step is at present non-rigorous. Hopefully it will be justified in the future.

Let us discuss how this works for the 3d Ising model: a lattice model with the
Hamiltonian $H = - \sum_{\langle x y \rangle} S_x S_y$ where $S_x = \pm 1$ are
spins on a cubic lattice, with the nearest-neighbor ferromagnetic interaction.

The ``3d Ising CFT'' is a CFT in $d=3$ describing the critical point of this model, and of any other
model in the same universality class. 
Just as the lattice Ising model, this CFT has a global $\mathbb{Z}_2$
invariance with all fields divided into $\mathbb{Z}_2$-even and
$\mathbb{Z}_2$-odd.\footnote{We have not included the notion of global
symmetry in the CFT axioms, but this extension is straightforward. It just
means that all fields transform in finite-dimensional irreducible
representations of a compact global symmetry group $G$, forming a direct
product with the conformal group. All correlators are invariant
tensors of $G,$ and the OPE respects this additional symmetry.} It
contains a $\mathbb{Z}_2$-odd scalar primary field denoted $\sigma (x)$, whose
correlators
\begin{equation}
  \langle \sigma (x_1) \sigma (x_2) \ldots \sigma (x_n) \rangle \label{ss}
\end{equation}
are interpreted as the 3d Ising model spin correlation functions
\begin{equation}
  \langle S_{x_1} S_{x_2} \ldots S_{x_n} \rangle \label{ssIsing}
\end{equation}
computed at the critical temperature $T = T_c$, at distances $| x_p - x_q |$
much larger than the lattice spacing. While $\langle \ldots \rangle$ in
{\eqref{ss}} is just a notation, in {\eqref{ssIsing}} it is a true average
with respect to the Gibbs distribution, in the thermodynamic limit. By Axiom
\ref{AxConf}, correlator {\eqref{ss}} is conformally invariant, and thus in
particular scale invariant, scale transformations being a part of the
conformal group. This means ($\Delta_{\sigma}$ is the scaling dimension of
$\sigma$):
\begin{equation}
  \langle \sigma (\lambda x_1) \sigma (\lambda x_2) \ldots \sigma (\lambda
  x_n) \rangle = \lambda^{- n \Delta_{\sigma}} \langle \sigma (x_1) \sigma
  (x_2) \ldots \sigma (x_n) \rangle .
\end{equation}
On the other hand, {\eqref{ssIsing}} clearly does not have such an exact scale
invariance, already because it is defined on a lattice. The precise statement
of agreement at large distances is\footnote{Equivalently, one can consider a
sequence of lattice models with a smaller and smaller lattice spacing $a$, and
take the limit $a \rightarrow 0$ while keeping $x_a$ fixed.}
\begin{equation}
  \lim_{| x_p - x_q | \rightarrow \infty} \frac{\langle S_{x_1} S_{x_2} \ldots
  S_{x_n} \rangle}{\langle \sigma (x_1) \sigma (x_2) \ldots \sigma (x_n)
  \rangle} = C^n, \label{long} 
\end{equation}
where $C$ is some constant, which is $n$-independent but non-universal (e.g.
it would change if we add next-to-nearest interactions to the lattice model,
which does not change the universality class).

Other 3d Ising CFT fields will correspond to other lattice-scale operators.
E.g. we can consider the product of two nearby spins (separated in an
arbitrary direction)
\begin{equation}
  E_x = S_x S_{x + 1} - \langle S_x S_{x + 1} \rangle,
\end{equation}
where $\langle S_x S_{x + 1} \rangle$ is subtracted so that $\langle E_x \rangle = 0$. The 3d Ising
CFT contains a $\mathbb{Z}_2$-even scalar primary $\varepsilon (x)$ whose
correlators describe long-distance limits of the $E_x$ correlators, similarly
to {\eqref{long}}.

More generally, we expect to have a CFT associated with every universality
class of continuous phase transitions. This CFT will share global symmetry
($\mathbb{Z}_2$, $O (N)$, etc) with the universality class, and its scaling
dimensions will determine the critical exponents. It has not been proven yet, starting from the lattice models or in any other way, that all these CFTs actually exist. This is the non-rigorous part of the CFT game.

CFT fields $A_i$ and their scaling dimensions $\Delta_i$ also have
counterparts in the RG approach to critical phenomena
{\cite{Wilson:1973jj}}.\footnote{On the other hand, the OPE coefficients
$\lambda_{i j k}$ do not feature prominently in the RG approach.} Namely, they
correspond to the eigenvectors and the eigenvalues of RG transformation
linearized near a fixed point describing a continuous phase transition. Fields
of scaling dimension $\Delta_i < d$ ($\Delta_i > d$) correspond to the
relevant (irrelevant) deformations of the fixed point. 
This dictionary is not needed for the
actual CFT calculations, but only for interpreting the results.

We expect that the above-mentioned fields $\sigma$ and $\varepsilon$ are the only two
relevant fields of the 3d Ising CFT. This follows from the experimental fact that the critical point of the 3d Ising model is in the same universality class as the liquid-vapor critical point, which is reached by tuning two parameters (pressure and temperature).

\section{Conformal bootstrap program}\label{bootstrap}

\subsection{Consistency}\label{sec:cons}

Conformal bootstrap program attempts to classify CFTs by classifying their
datasets. That this may be possible was first suggested by Polyakov \cite{Polyakov:1974gs}.

We call a dataset $\mathcal{D} = \{ \Delta_i, \lambda_{i j k} \}$ ``consistent'' if
it is a dataset of some CFT: $\mathcal{D} = \text{Data} (\mathcal{T})$. Ideally, we
would like to have a list of all consistent data sets:\footnote{We are not
giving full details necessary to make this statement precise. One important
subclass of CFTs are ``local CFTs'', which roughly correspond to critical
points of lattice models with finite-range interactions. It is expected that
most local CFTs are isolated. One exception are CFTs with ``exactly marginal''
fields of dimension $\Delta = d$, which form finite-dimensional continuous
families. A folk conjecture says that exactly marginal fields in $d \geqslant
3$ require supersymmetry, which makes this exception non-generic.}
\begin{equation}
\{ \text{Data} (\mathcal{T}_1), \text{Data} (\mathcal{T}_2), \ldots \}, 
\end{equation}
but it is not currently known how to generate such a list. The following question is
less ambitious but still very interesting:
\begin{equation}
  \text{\textit{Q1: Given a trial dataset $\mathcal{D}$, decide if it is
  inconsistent}.} \label{Q1}
\end{equation}
It turns out that this has an algorithmic answer. This will allow progress on
classification by ruling out inconsistent data sets (rather than by
constructing consistent ones).

The idea is straightforward: given a trial dataset $\mathcal{D}= \{ \Delta_i,
\lambda_{i j k} \}$, we will try to construct all correlators, looking for
some inconsistency with the axioms.

The first step is to construct the 2pt and 3pt correlators. \ These are simply
given by explicit formulas from Axiom \ref{AxSimple}(d) and {\eqref{3pt}} with
$c_{i j k} = \lambda_{i j k}$. So far no room for inconsistency.

\begin{figure}[h]
	\centering
  {\includegraphics[width=0.3\linewidth]{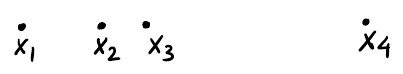}}
  \caption{\label{Ex4pt}The 4pt correlator for this configuration of points can be reduced to 3pt
  correlators using the OPE {\eqref{OPEeq}} with $(x, y)$ being one of the
  following pairs: $(x_1, x_2)$, $(x_2, x_3)$, $(x_3, x_2)$, $(x_4, x_3)$. }
\end{figure}

Then we proceed to construct the 4pt correlators. For this we consider the OPE
series reducing them to the 3pt correlators. All information needed to write
down these series is contained in $\Delta_i$ and $\lambda_{i j k}$. But now we
need to check a couple of things. First, do these series converge where Axiom
\ref{AxOPE} says they should? For this, the trial OPE coefficients $\lambda_{i
j k}$ should not grow too fast as a function of $k$ for fixed $i$, $j$. The
required growth condition can be shown to take a relatively simple form:
\begin{equation}
  \sum_{k = 0}^{\infty} (4 \rho)^{\Delta_k} \lambda^2_{i j k} < \infty \qquad
  \forall \rho < 1. \label{growth}
\end{equation}
Second, there are several ways to reduce a 4pt correlator to 3pt correlators via
the OPE, and they all should agree in the overlapping regions of convergence.
See Fig. \ref{Ex4pt} for an example. This condition is called ``crossing'',
and it is not automatically satisfied.\footnote{4pt crossing constraints were first discussed in Refs.~\cite{Ferrara:1973yt,Polyakov:1974gs}. The word ``crossing'' comes from an analogy with relativistic Quantum Field Theory. There, the $2\to 2$ scattering amplitude $\mathcal{M}(p_1,p_2 \to p_3,p_4)$ is invariant under ``crossing transformations", when one incoming particle is moved (``crosses'') into the group of outgoing particles, while one outgoing particle crosses in the opposite direction.} Assuming that it also holds, we can
define the 4pt correlators as the sum of OPE series.

We then proceed to higher $n$-point correlators. Similarly to $n = 4$, they are
reduced to $(n - 1)$-point correlators via the OPE, and we need to check
convergence and crossing. It turns out that crossing for $n \geqslant 5$ is
automatically satisfied once we impose crossing for all 4pt correlators. On the
other hand, the general convergence condition is stronger than
{\eqref{growth}}, and it can be expressed as follows. Consider an infinite
matrix consisting of OPE coefficients $\lambda_{i j k}$ with a fixed $j$ and
arbitrary $i, k$:
\begin{equation}
  (M^{(j)})_{i k} = \lambda_{i j k} .
\end{equation}
Then any such $M^{(j)}$, viewed as an operator from $i$-indexed sequences to
$k$-indexed sequences, should be bounded with respect to a certain weighted
$\ell_2$ norm (with $\Delta$-depending weights).\footnote{This is related to
something called `radial quantization', which we do not describe in this text.
This convergence condition for higher $n$-point corelators has not been
discussed in detail in the literature.} Eq.~{\eqref{growth}} follows
from this general condition when we apply the operator to a sequence
consisting of a single nonzero element, and demand that the result have a
finite norm.

To summarize, consistent datasets are those which satisfy the general
convergence and the 4pt crossing conditions. The convergence condition is the
less interesting of the two. Below we will focus on the 4pt crossing, which
will allow us to put constraints on the fields of low scaling dimension.

\subsection{Conformal blocks and 4pt crossing}\label{n4cross}

Here we will describe how to put 4pt crossing constraint into a more explicit
form, by expanding 4pt correlators in a basis of special functions called
conformal blocks.

Recall that conformally invariant 4pt correlators have form {\eqref{4pt}}. When
we compute the 4pt correlator in the r.h.s. of {\eqref{4pt}} using the OPE, we
should get something consistent with this formula. Let us see how this
happens. Applying the OPE to the first pair of fields, we get an expression of
the form:
\begin{equation}
  \langle A_i (x_1) A_j (x_2) A_k (x_3) A_l (x_4) \rangle = \sum_m
  \frac{\lambda_{i j m}}{x_{12}^{\Delta_i + \Delta_j - \Delta_m}} [\langle A_m
  (x_1) A_k (x_3) A_l (x_4) \rangle + \cdots], \label{4ptOPE}
\end{equation}
where {...} denotes terms proportional to $s_{i j m}^{(r)}$ times
derivatives acting on the 3pt correlator $\langle A_m A_k A_l \rangle$, which is
in turn given by $\lambda_{m k l}$ times an $x$-dependent function which can
be read off from {\eqref{3pt}}. It can be shown that by doing all derivatives
and infinite sums over $r$, the r.h.s. of Eq.~{\eqref{4ptOPE}} takes the form: \
\begin{equation}
  \left( \frac{x_{24}}{x_{14}} \right)^{\Delta_i - \Delta_j} \left(
  \frac{x_{14}}{x_{13}} \right)^{\Delta_k - \Delta_l}
  \frac{1}{x_{12}^{\Delta_i + \Delta_j} x_{34}^{\Delta_k + \Delta_l}} \sum_m
  \lambda_{i j m} \lambda_{m k l} G_{\Delta_m} (u, v) . \label{4ptOPE1}
\end{equation}
The functions $G_{\Delta_m} (u, v)$ appearing here are called `conformal
block'. These functions are fixed by conformal symmetry. They depends on the
exchanged scaling dimension $\Delta_m$, and on the space dimension
$d$.\footnote{They also depend on the external dimension differences $\Delta_i
- \Delta_j$, $\Delta_k - \Delta_l$ but we will omit this from the notation. In
a full treatment involving spinning fields, the conformal blocks also depend
on the spin of the fields. } Notably, they do not depend on the OPE
coefficients $\lambda_{i j k}$ whose product appears as a prefactor in
{\eqref{4ptOPE1}}.

Theory of conformal blocks is huge and it's not possible to do it justice in
this text. It has connections to representation theory, orthogonal
polynomials, and integrable quantum mechanics. There are no fully general
closed form expressions of conformal blocks in terms of the classical special
functions. Fortunately, they admit rapidly convergent power series expansions
which allow efficient numerical evaluation. This is what is used in practical
applications.

The conformal block is simple only for the exchanged unit field: $A_m = A_0 =
1$, when we have:
\begin{equation}
  G_0 (u, v) = 1, \quad \lambda_{i j 0} = \delta_{i j}, \quad \lambda_{0 k l}
  = \delta_{k l}, \label{G0}
\end{equation}
where we also gave the OPE coefficients for this case (see footnote
\ref{lNote}).

Comparing Eq.~{\eqref{4ptOPE1}} with {\eqref{4pt}} we see that they are consistent if we identify:
\begin{equation}
  g_{i j k l} (u, v) = \sum_m \lambda_{i j m} \lambda_{m k l} G_{\Delta_m} (u,
  v) . \label{4ptCB}
\end{equation}
This gives a compact formula to compute the 4pt correlators in terms of the CFT
data. We can also obtain a compact expression for the 4pt crossing
constraints, by substituting Eq.~{\eqref{4ptCB}} into
{\eqref{4ptcross}}:
\begin{equation}
  u^{- \frac{\Delta_i + \Delta_j}{2}} \sum_m \lambda_{i j m} \lambda_{m k l}
  G_{\Delta_m} (u, v) = v^{- \frac{\Delta_k + \Delta_j}{2}} \sum_m \lambda_{k
  j m} \lambda_{m i l} G_{\Delta_m} (v, u) . \label{4ptcr}
\end{equation}
As Eq.~{\eqref{4ptcross}}, this corresponds to the permutation $x_1
\leftrightarrow x_3$. Constraints corresponding to other permutations take a rather similar form. They should
also be considered, although we will not
discuss them here explicitly.

Now, we can test a trial dataset $\mathcal{D}$ for consistency, by checking Eq.
{\eqref{4ptcr}} for all possible choices of $i, j, k, l$, in the region of
overlapping convergence. This region is not empty. E.g., let us fix points
$x_{1, 3, 4}$ so that $x_4$ is far away from $x_{1, 3}$. Then the l.h.s.
should converge within the set $\{ x_2 : x_{12} < x_{13} \}$, and the r.h.s.
in $\{ x_2 : x_{23} < x_{13} \}$. These two balls have a nontrivial overlap.

\subsection{Partially specified datasets}

In sections \ref{sec:cons} and \ref{n4cross}, we gave an answer to the 
consistency question {\eqref{Q1}}. Unfortunately, the described procedure
is not by itself practically useful, since it assumes that the trial dataset
$\mathcal{D}$ is fully specified, which includes infinitely many parameters (all scaling
dimensions and OPE coefficients). To correct for this, let us define the
notion of a ``partially specified trial dataset'', which is a list
$\mathcal{L}$ of finitely many assumptions on scaling dimensions and OPE
coefficients. We say that $\mathcal{L}$ is consistent if there is at least one
CFT $\mathcal{T}$ whose dataset $\text{Data} (\mathcal{T})$ satisfies the assumptions. The
following is then a more practical version of question {\eqref{Q1}}:
\begin{equation}
  \text{\textit{Q2: Given a partially specified trial dataset $\mathcal{L}$,
  decide if it is inconsistent}.} \label{Q2}
\end{equation}
Although this looks like a much harder question than {\eqref{Q1}}, it turns
out that this question can also be answered, based on Eq.~{\eqref{4ptcr}},
using numerical algorithms. This was first shown by Rattazzi, Tonni, Vichi and the author {\cite{Rattazzi:2008pe}}
and led to the rapid development of the numerical conformal bootstrap in the
last 10 years. We will explain how this work on an example in the next
section.

\section{Example: constraining the 3d Ising CFT}\label{ex}

Let us fix two real numbers $\Delta_1, \Delta_2$ in the interval $[\frac12, 3]$, and consider the
following list of assumptions $\mathcal{L}=\mathcal{L} (\Delta_1, \Delta_2)$
about a 3d CFT:
\begin{itemize}
  \item $\mathbb{Z}_2$ global symmetry;
  \item there is one field which is $\mathbb{Z}_2$-odd, one which is
  $\mathbb{Z}_2$-even, and they have scaling dimensions $\Delta_1$ and
  $\Delta_2$;
  \item all other fields have scaling dimensions $\Delta_i \geqslant 3$ i.e.
  are irrelevant.
\end{itemize}
As discussed in section \ref{interpretation}, the 3d Ising CFT satisfies
$\mathcal{L} (\Delta_{\sigma}, \Delta_{\varepsilon})$. Our strategy will be to
exclude a large part of the $(\Delta_1, \Delta_2)$-plane by showing
that $\mathcal{L} (\Delta_1, \Delta_2)$ is inconsistent there. This will imply
that the scaling dimensions of the 3d Ising CFT must belong to the remaining
part of the plane.

\subsection{One crossing constraint}\label{one-constr}

Consider first the 4pt crossing for $\langle A_1 A_1 A_1 A_1 \rangle$. Putting
$i = j = k = l = 1$ in {\eqref{4ptcr}}, we obtain:
\begin{equation}
  u^{- \Delta_1} \sum_m p_m G_{\Delta_m} (u, v) = v^{- \Delta_1} \sum_m p_m
  G_{\Delta_m} (v, u), \qquad p_m = \lambda^2_{11 m} \geqslant 0
  \label{4ptcr1111} .
\end{equation}
We know that $p_0 = G_0 = 1$ (see {\eqref{G0}}), so isolating those terms we
write this as
\begin{eqnarray}
  & h (u, v) \equiv v^{- \Delta_1} - u^{- \Delta_1} = \sum_{m = 2}^{\infty}
  p_m F^{}_{\Delta_m} (u, v), &  \label{pmeq}\\
  & F^{}_{\Delta} (u, v) := u^{- \Delta_1} G_{\Delta_{}} (u, v) - v^{-
  \Delta_1} G_{\Delta_{}} (v, u) . &  \nonumber
\end{eqnarray}
Note that $F_{\Delta}$ also depends on $\Delta_1$. The sum starts from $m = 2$
because $\lambda_{111} = 0$ for the $\mathbb{Z}_2$-odd $A_1$.

Geometrically, {\eqref{pmeq}} means that $h$, viewed as a vector in a space of
two-variable functions, belongs to a convex cone $\mathcal{C}_{}$ generated by
vectors $F^{}_{\Delta_2}$ and $F_{\Delta}$ with $\Delta \geqslant 3$. We
include all $F_{\Delta}$ with $\Delta \geqslant 3$ as generators of the cone
since we don't know the exact values of $\Delta_m$ for $m \geqslant 3$, but
only that $\Delta_m \geqslant 3$. Denote by $\mathcal{C}_{}^{\ast}$ the dual
convex cone, which is the set of all linear functionals $\alpha$ which are
positive on all vectors generating the cone:
\begin{equation}
  \alpha [F_{\Delta_2}] \geqslant 0, \quad \alpha [F_{\Delta}] \geqslant 0
  \quad \forall \Delta \geqslant 3. \label{C*}
\end{equation}
Suppose that there exists a functional $\alpha_0 \in \mathcal{C}_{}^{\ast}$
such that
\begin{equation}
  \alpha_0 [h] < 0. \label{hneg}
\end{equation}
Then by acting with $\alpha_0$ on Eq.~{\eqref{pmeq}} we get a contradiction.
So, this equation cannot be satisfied for any nonnegative $p_m$. This is how
one shows that the assumption $\mathcal{L} (\Delta_1, \Delta_2)$ is
inconsistent: by exhibiting a functional $\alpha_0$ which satisfies
{\eqref{C*}} and {\eqref{hneg}}.

Numerically, one works with a finite dimensional space of functionals
$\mathcal{A} (\Lambda)$ which are finite sums of partial derivatives
at a particular point:
\begin{equation}
  \alpha [f] = \sum_{m + n \leqslant \Lambda} \alpha_{m, n} \partial_u^m
  \partial_v^n f (u_0, v_0), \label{Nder}
\end{equation}
where $\Lambda$ is a parameter, to be taken as large as possible to have the
maximal constraining powers (within the available computer resources). One
then minimizes $\alpha [h]$ over all $\alpha \in \mathcal{C}_{}^{\ast} \cap
\mathcal{A} (\Lambda)$, looking for a functional satisfying
{\eqref{hneg}}. This is a convex optimization problem (continuous linear programming), which can be solved by
efficient numerical algorithms. If the minimum is negative, then we ruled out
$\mathcal{L} (\Delta_1, \Delta_2)$. If it is positive, and cannot be made
negative by increasing $\Lambda$, this would mean that $\mathcal{L} (\Delta_1,
\Delta_2)$ is consistent with crossing for $\langle A_1 A_1 A_1 A_1 \rangle$.

With this procedure, Ref. {\cite{ElShowk:2012ht}} showed that the constraint
$\mathcal{L} (\Delta_1, \Delta_2)$ is inconsistent in a significant portion of
parameter space. Invoking an extra and so far unproven assumption, that the 3d
Ising CFT lies at a singular boundary point of the consistent region (the so
called ``kink''), Refs. {\cite{ElShowk:2012ht,El-Showk:2014dwa}}
gave the first conformal bootstrap determination of $\Delta_{\sigma},
\Delta_{\varepsilon}$. Subsequent work has shown that the kink assumption is
unnecessary, provided that one includes crossing constraints for the 4pt
correlators $\langle A_1 A_1 A_2 A_2 \rangle$, $\langle A_2 A_2 A_2 A_2
\rangle$. We will now explain briefly how this was done.

\subsection{Several crossing constraints}

\begin{figure}[h]
	\centering
{\includegraphics[width=8.65730355503083cm,height=5.25567361930998cm]{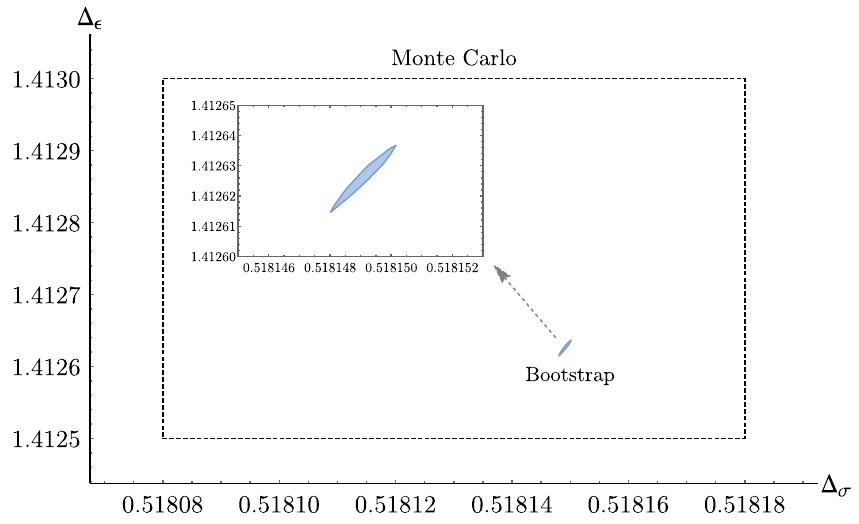}}
  \caption{\label{island}The bootstrap island to which the 3d Ising CFT must
  belong {\cite{Kos:2016ysd}}. Also shown is a Monte Carlo result \cite{Hasenbusch2010} for the same
  scaling dimensions. This plot used $\Lambda = 43$ in {\eqref{Nder}}.} %Future work will shrink the bootstrap island further by increasing $\Lambda$.
\end{figure}

To increase the constraining power, a natural idea is to include crossing
constraints for the other 4pt correlators of fields $A_1$ and $A_2$. While
$\langle A_2 A_2 A_2 A_2 \rangle$ is completely analogous to $\langle A_1 A_1
A_1 A_1 \rangle$, one encounters a crucial difference when analyzing $\langle
A_1 A_1 A_2 A_2 \rangle$. Namely, its conformal block expansion involves
products of two different OPE coefficients $\lambda_{11 m} \lambda_{22 m}$.
These products are not necessarily positive, because $\lambda_{i j k}$ may
have either sign. On the other hand, positivity of the coefficients $p_m =
\lambda^2_{11 m}$ played a crucial role in making the minimization problem of
section \ref{one-constr} convex. To overcome this obstacle, one analyzes all
three correlators together, and considers the matrix
\begin{equation} P_m = \left(\begin{array}{cc}
     \lambda_{11 m}^2 & \lambda_{11 m} \lambda_{22 m}\\
     \lambda_{11 m} \lambda_{22 m} & \lambda_{22 m}^2
   \end{array}\right) . 
   \end{equation}
Crucially, this matrix is positive semidefinite: $P_m \succcurlyeq 0$. This
condition is convex, and provides a good substitute for the simple positivity
in the bootstrap problems involving multiple correlators. The resulting
problem is that of continuous semidefinite programming, and it can still be attacked by efficient numerical algorithms. This was
realized and carried out in Refs. {\cite{Kos:2014bka,Simmons-Duffin:2015qma,Kos:2016ysd}} which found a consistent
``island'' near $\Delta_1 \approx 0.5181489 (10)$, $\Delta_2 \approx 1.412625
(10)$. The 3d Ising CFT point $(\Delta_\sigma,\Delta_\varepsilon)$ must live somewhere in this tiny island  (Fig. \ref{island}).

The scaling dimensions $\Delta_\sigma, \Delta_\varepsilon$ determine the main critical exponents of the 3d Ising model $\alpha, \beta, \gamma,\delta,\eta,\nu$. In what follows we will focus on $\eta$ and $\nu$, given by
%\begin{equation}
%\eta = 2 \Delta_{\sigma} - 1\,,\qquad \nu
%= 1 / (d - \Delta_{\varepsilon})\,.\label{etanu}
%\end{equation}
\begin{eqnarray}
\eta &=& 2 \Delta_{\sigma} - 1\,,\label{eta}\\ 
\nu &=& 1 / (d - \Delta_{\varepsilon})\,.\label{nu}
\end{eqnarray}
Eq.~\eqref{nu} deserves a comment, because it expresses an off-critical quantity (the exponent $\nu$ describing behavior of the correlation length close to the critical point) via a critical theory parameter $\Delta_{\varepsilon}$. 
This is an example of how CFT can make predictions about small deviations from the critical theory, which arise at short distances from relevant perturbations , and at large distances from the irrelevant ones. Such predictions are done via a technique called ``conformal perturbation theory,'' which we have not explained. Eq. \eqref{omega} below is another simple example.

Another important quantity is the ``correction to scaling'' exponent $\omega$. 
It appears in the rate $\sim1/r^\omega$ at which the limit in \eqref{long} is achieved, assuming that all distances $|x_p-x_q|\sim r$ are of the same order. It also appears in the subleading singularities of all quantities exhibiting powerlaw behavior near the critical point (e.g. the specific heat). 
While describing deviations from criticality, $\omega$ like $\nu$ can be expressed in terms of a purely critical parameter:
\begin{equation}
	\label{omega}
	\omega=\Delta_3-d\,,
	\end{equation}
where $\Delta_3$ is the scaling dimension of the leading irrelevant $\mathbb{Z}_2$-even scalar operator. The conformal bootstrap determines $\Delta_3$ (and hence $\omega$) by scanning the island in Fig.~\ref{island} and reconstructing the spectrum which provides a solution to the 4pt crossing \cite{Simmons-Duffin:2016wlq}. 

In Table \ref{tab:comp} we report the values of the critical exponents $\nu$, $\eta$, $\omega$ according to the conformal bootstrap, Monte Carlo simulations and RG calculations. We also include some experimental measurements of $\nu$ and $\eta$.\footnote{Since $\omega$ parametrizes subleading powers, it is harder to measure, and we are not aware of any published result.}
The conformal bootstrap predictions are the most precise, and they are in a good agreement with the Monte Carlo and RG. There is also reasonable agreement between the theory and the experiment, although the experimental accuracy is not amazing.

\begin{table}[h]\centering
	\begin{tabular}{@{}llllll@{}}
		\toprule
		 Ref& Year & Method/Experiment & $\quad\nu$ & $\quad\eta$ & $\quad\omega$\\
		 \midrule
\cite{Kos:2016ysd,Simmons-Duffin:2016wlq} & 2016 & Conformal bootstrap   & 0.629971(4) & 0.036298(2) & 0.82968(23) \\
\cite{Hasenbusch2010} & 2010 & Monte Carlo & 0.63002(10)  & 0.03627(10)   & 0.832(6)\\
\cite{Guida:1998bx} & 1998 & RG & 0.6304(13) & 0.0335(25) & 0.799(11)\\
\midrule
\cite{PhysRevB.40.4696}  & 1989 & Binary fluid & 0.628(8) & 0.0300(15) & \\
\cite{Fluids} &2009 & Binary fluid & 0.629(3) & 0.032(13) &\\
\cite{doi:10.1002/andp.19945060102} & 1994 & Binary mixture & 0.623(13) & 0.039(4) & \\
\cite{Sullivan_2000} & 2000 & Liquid-vapor&  0.62(3)   & & \\
\cite{PhysRevB.58.12038} & 1998 & Liquid-vapor && 0.042(6) &\\
\cite{PhysRevB.35.4823} & 1987 & Uniaxial antiferromagnet & 0.64(1) & \\
		\bottomrule\\
	\end{tabular}
	\caption{\label{tab:comp} Some representative theoretical and experimental determinations of the 3d Ising critical exponents. See \cite{Pelissetto:2000ek}, Section 3.2, for more references. }
\end{table}

%\begin{table}[h]\centering
%	\begin{tabular}{@{}llllll@{}}
%		\toprule
%		Exponent & Value & Method/Experiment & Year & Ref \\
%		\midrule
%		$\nu$  &0.629971(4) &Conformal bootstrap  & 2016 & \cite{Kos:2016ysd}  \\
%		& 0.63002(10)  & Monte Carlo & 2010 & \cite{Hasenbusch2010}\\
%		& 0.6304(13)   & RG & 1998 & \cite{Guida:1998bx}\\
%		
%		
%		$\eta$ &  0.036298(2)  & Conformal bootstrap& 2016 &  \cite{Kos:2016ysd}  \\
%		& 0.03627(10)  & Monte Carlo & 2010 & \cite{Hasenbusch2010}\\
%		& 0.0335(25)   & RG & 1998 & \cite{Guida:1998bx}\\
%		\bottomrule\\
%	\end{tabular}
%\caption{\label{tab:comp}Caption. We also include...}
%\end{table}

\section{Conclusions}

Conformal bootstrap calculations provide predictions for observable physical
quantities from the CFT axioms. Agreement of these
predictions with alternative theoretical determinations and the experiment
increase our belief in the validity of the axioms.

Feynman {\cite{Feynman}} called the Gibbs distribution the ``summit of
statistical mechanics'', the entire subject being either the ``climb-up'' to
derive it, or the ``slide-down'' when it is applied. Echoing Feynman, we may call
the CFT a summit of the theory of critical phenomena, the conformal
bootstrap being the way to slide down. To climb up would be to prove the validity of the interpretation of the CFT axioms described in Section \ref{interpretation}. Unfortunately, relatively little
rigorous work has been done in the way of climbing up.\footnote{The last 20
years, starting with Smirnov {\cite{SMIRNOV}}, have seen significant progress
in showing rigorously conformal invariance of specific 2d models. This program
is still far from establishing conformal invariance of a generic critical
theory, let alone the full scope of the CFT axioms, such as the existence of a
complete set of local operators and the OPE. We are also not aware of any relevant
mathematical work in $d \geqslant 3$.} One should also not forget a second major peak in
the same mountain range: the Renormalization Group.

\section*{Acknowledgements}

This article is based on the talk at the mathematical physics workshop ``Inhomogeneous
Random Systems'' (Institut Curie, Paris, January 28, 2020). I am grateful Ellen Saada, Fran{\c
c}ois Dunlop and Alessandro Giuliani for the organization and the invitation
to speak. I am also grateful to Jacques Villain for the invitation to write this article, careful reading of the draft, and many suggestions on how to improve the presentation. SR is partly supported by the Simons Foundation grant 488655 (Simons
Collaboration on the Nonperturbative Bootstrap), and by Mitsubishi Heavy
Industries as an ENS-MHI Chair holder.

\small

%\bibliographystyle{utphys}
%\nocite{*}
%\bibliography{references}

\providecommand{\href}[2]{#2}\begingroup\raggedright\endgroup

\end{document}